\documentclass[12pt]{article}
\usepackage{graphics}
\input epsf

\newcommand{\be}{\begin{eqnarray}}
\newcommand{\ben}{\begin{eqnarray}\nonumber}
\newcommand{\ee}{\end{eqnarray}}

\begin{document}
\title{
\vskip 1cm
Nuclear Physics in a Susy Universe
}
\author{L. Clavelli\footnote{lclavell@bama.ua.edu}$\;$ and I. Perevalova
\footnote{perev001@bama.ua.edu}\\
Department of Physics and Astronomy\\
University of Alabama\\
Tuscaloosa AL 35487\\ }
\date{July 12, 2010}
\maketitle
\begin{abstract}
     We refine a previous zeroth order analysis of the nuclear properties of a supersymmetric (susy) universe with standard model particle content plus degenerate susy partners.  No assumptions are made concerning the Higgs structure except we assume that the degenerate fermion/sfermion masses are non-zero.  This alternate universe has been dubbed Susyria and it has been
proposed that such a world may exist with zero vacuum energy in the string landscape.
\end{abstract}


\section{\bf Introduction: Susyria Revisited}
\setcounter{equation}{0}
    Since the observation \cite{Weinberg} that the evolution of intelligent life requires a low vacuum energy and the subsequent consistent experimental result,  it has been of interest to ask what kind of alternate universes might have a physics consistent with advanced life forms (ALF).  Current ideas based on the string landscape \cite{Susskind, Bousso, Linde, Giddings}suggest that there are a huge number of local minima in the effective potential that are consistent with the rise of ALF and many more minima without the possibility of ALF.  Ideally, we should one day understand the existence of a universe with our specific vacuum energy and the history
leading to it at the end of the inflationary era.  It is well known that small deformations of some individual physical parameters \cite{Barr, Donoghue} would lead to a hopelessly toxic universe but it is not ruled out that a simultaneous deformation of multiple parameters could result in a complete body of physical law friendly to ALF.  An example is the weakless universe \cite{weakless}, whose gauge group lacks the SU(2) of our standard model.  In that paper it was shown that other parameters could be adjusted to allow such a universe to produce and distribute the heavy elements that seem to be essential to life.  Of course, ref.\,\cite{weakless} did not explore any subsequent requirements for the appearance of ALF and it has been pointed out \cite{CW} that many obstacles remain before the weakless universe could be proven hospitable to life.  Thus no confirmed example has been discovered of a substantially different physical law that is clearly consistent with ALF.  

   We would assume that transitions among local (or absolute) minima in the string landscape would, as in Lagrangian Higgs models, preserve the number of degrees of freedom while readjusting masses.  Thus all possible universes connected to ours by a possible vacuum decay should contain at least the equivalent of our standard model particles.  It is, therefore, not clear how or whether the anti-DeSitter universe of the ADS/CFT correspondence \cite{ADS} and the metastable models of ref.\,\cite{ISS} are related to our universe by a possible phase transition.   

    We have previously suggested \cite{future} that a prime example of an alternate universe would be Susyria, an exactly supersymmetric universe similar to the Minimal Supersymmetric Standard Model (MSSM) with soft parameters turned off and, presumably, with an extended Higgs structure \cite{Xhiggs} so that particles would have non-zero masses in the susy phase.
The last requirement would be essential if electromagnetic bound states (atoms) are to occur.  Thus we explore what minimal properties Susyria would have to have to allow the rise of supersymmetric life.  This is not to say that other sterile susy universes (Susalia) do not exist.
     
   A natural property of Susyria is a vanishing vacuum energy so that the jump into it from our positive vacuum energy universe is much more probable than the opposite transition from the exact susy state to our
broken susy universe.  There may, however, be an anthropic argument that the universe could not have fallen directly into the susy phase at the end of the inflationary era.  The de-tuning of the triple alpha coincidence as discussed below would probably have precluded the formation of long-lived stars essential to ALF.  On the other hand, we ask whether, given a prior evolution of stars and planets in the broken susy phase, ALF could re-establish themselves following a transition to the supersymmetric phase.

   We have shown previously that, given a small nuclear bound state of degenerate nucleons and snucleons, atoms will form and both ionic and covalent molecular bound states will exist \cite{CL,CS}.  It remains to determine which nuclear isotopes are present in a susy world.  A zeroth order answer to this question was given in ref.\,\cite{future} and the purpose of this note is to refine that analysis.

\section{Elements of Snuclear Physics}

   As in ref.\,\cite{future} we begin with the semi-empirical mass formula for nuclear states of $Z$ protons, $N$ neutrons, and a total of $A$ nucleons.
The masses of standard nuclei can be fit to great precision by a sum of a
volume coefficient, $a_V$, a surface coefficient, $a_S$, a Coulomb coefficient, $a_C$, an asymmetry coefficient, $a_A$, and an alternating
term proportional to $\delta$ encoding the preference for even numbers of neutrons and protons.

\ben
    M(Z,A) = m_N\;N + m_P\;Z - a_V\;A + a_S\;A^{2/3}&+&a_C\;\frac{Z^2}{A^{1/3}}
          +  a_A\;\frac{(N-Z)^2}{A}\\ 
&+& \delta\; \frac{\cos(\pi\,Z) \cos^2(\pi\,A/2)}{\sqrt A}
\label{M1(Z,A)}
\ee 
An excellent fit to hundreds of nuclear masses is defined by the coefficients
\be\nonumber
     &a_V = & 15.67\, {\displaystyle MeV}\\ \nonumber
     &a_S = & 17.23\, {\displaystyle MeV}\\ 
     &a_C = & 0.714\, {\displaystyle MeV}\\ \nonumber
     &a_A = & 23.3\, {\displaystyle MeV}\\ \nonumber
     &\delta =& -11.5\,  {\displaystyle MeV}
\ee
If a nucleus is overtaken by a susy bubble, pairs of identical fermions trapped in high energy levels by the Pauli Principle will convert in pairs to
their degenerate bosonic partners due to gluino exchange. Then, being insensitive to the Pauli Principle, these will drop into the ground state
emitting a gamma ray burst.  The analogous process via photino exchange in a degenerate electron gas has been treated in ref.\,\cite{CP}.  If the nucleus is surrounded by a degenerate electron gas as in a white dwarf star, only the resulting photons below the Fermi energy of the gas will readily escape.
This is in qualitative agreement with the energy of the gamma ray burst photons.  Afterwards, nuclear reactions among nuclei with bosonic constituents will provide new sources of energy which could extend stellar lifetimes or facilitate stellar explosions \cite{BC}.   

The final state nuclei will be essentially free of the Pauli Principle.
This is one of the most prominent features of a susy nucleus.  We can ask how the semi-empirical mass formula might be changed by the effective absence of the Pauli Principle.  In ref.\,\cite{future} we assumed that the asymmetry term
was totally due to the exclusion principle and would therefore be absent in a susy nucleus.  We also set the $\delta$ term to its minimum as in an even-even nucleus. To lowest order we would expect the coupling strengths to remain approximately unaffected by the susy transition.  Without the asymmetry term, the resulting 
nuclear formula implied that a very large neutron number would be required to stabilize nuclei of medium to high atomic number.  For example, the lightest isotope of oxygen would then be $O^{209}$.  
In the spirit of pursuing the possibilitties that would maximize the
likelihood of the evolution of ALF in an alternate universe as was done in the weakless example, we assume in this paper that the nuclear masses in our world result from some potential energy terms and the $\delta$ term plus a kinetic energy piece given, at least for moderate to large atomic weight, by the Fermi gas model. 
That is, we write
\ben
    M(Z,A) = m_N\;N + m_P\;Z - {\tilde a}_V\;A &+& a_S\;A^{2/3} + a_C\;\frac{Z^2}{A^{1/3}}
          +  {\tilde a}_A\;\frac{(N-Z)^2}{A}\\ 
&+& \delta\; \frac{\cos(\pi\,Z) \cos^2(\pi\,A/2).}{\sqrt A} + E_P \quad .
\label{M2(Z,A)}
\ee 
The Pauli energy in the Fermi gas model is
\ben
   E_P &=& \frac{3\,A\, (\hbar\,c)^2}{20 M_N\,R_0^2}(9\pi/8)^{2/3} \cdot
         \left[(2 Z/A)^{5/3} + (2(A-Z)/A)^{5/3} \right]\\ 
       &\approx& 20.0 {\displaystyle MeV}\;\frac{A}{2}\; \left[(2 Z/A)^{5/3} + (2(A-Z)/A)^{5/3} \right]
\label{E_P}
\ee
where $R_0$ is the nucleon size parameter $1.2\,fm$.  The Pauli energy is minimized at fixed $A$ by equal numbers of neutrons and protons.  Expanding $E_P$ about this symmetry point we find an anti-binding term proportional to $A$ and a contribution to the asymmetry term with higher order terms negligible near the minimum.
\be
    E_P = A \cdot 20.0 {\displaystyle MeV} + \frac{(Z-N)^2}{A} \cdot 11.5 {\displaystyle MeV} + ... \quad . 
\ee
The consistency between eq.\,\ref{M1(Z,A)} and eq.\,\ref{M2(Z,A)} requires
\be
     \tilde{a}_V &=& a_V + 20.0\, {\displaystyle MeV} = 35.6\, {\displaystyle MeV} \\
     \tilde{a}_A &=& a_A - 11.1\, {\displaystyle MeV} = 12.2\, {\displaystyle MeV} 
\ee
The terms in eq.\,\ref{M2(Z,A)} that are clearly due to the Pauli Principle are the $\delta$ term above its minimum and the $E_P$ term.  If we discard these the suggested ground state mass for a susy nucleus of atomic number $Z$ and atomic weight $A$ is
\be
    M_s(Z,A) = m_N\,N + m_P\,Z - {\tilde a}_V\,A + a_S\,A^{2/3} + a_C\,\frac{Z^2}{A^{1/3}}
          +  {\tilde a}_A\,\frac{(N-Z)^2}{A} -\frac{11.5\,{\displaystyle MeV}}{A^{1/2}}\quad .
\label{M3(Z,A)}
\ee 
Taking the Coulomb coefficient, $a_C$ to be the same as in the standard model is equivalent to the assumption that the nuclear radius is still approximately $R_0\,A^{1/3}$.  This, in turn, requires that the susy nucleons have approximately the same radius as in our world and the susy nucleons are subject to a hard core potential.  The nucleon radius is a result of QCD and should be approximately the same for scalar nucleons in a susy universe.  Hard
core nuclear potentials have long been found necessary
even in helium,  where all the nucleons are in the ground state unaffected by the Pauli Principle
\cite{JainSrivastava}.
Bosonic nucleons will still be subject to the repulsive effect of the scalar $\sigma$ exchange while binding will come not from pion exchange but from heavier resonances.  Clearly, our assumption here of the same dependence of nuclear radius on atomic weight for scalar as for fermionic nucleons is only
a preliminary approximation tolerable for now only because of the smallness of the Coulomb term.  

Accepting the mass formula of eq.\,\ref{M3(Z,A)} for susy nuclei, stability against beta decay requires:
\be
    M_s(Z,A) < M_s(Z+1,A) + m_e
\ee
or 
\be
    2\,Z + 1 > A\,\eta
\ee
where
\be
       \eta = \frac{4\,\tilde{a}_A + m_N - m_P - m_e}{4\,\tilde{a}_A + a_C\,A^{2/3}} \quad .
\ee
Similarly stability against K capture, which also implies stability against $\beta^{+}$ decay,
requires:
\be
       2\,Z - 1 <  A\,\eta \quad .
\ee

Thus the nucleus $(Z,A)$ is stable against weak interactions if
\be
      -1 + A\,\eta < 2\,Z < 1 + A\,\eta \quad .
\label{betaconstraints}
\ee
Stability against $\alpha$ decays requires
\be
      M_s(Z,A) < M_s(Z-2,A-4) + m_\alpha\quad .
\label{alphadecay}
\ee
The mass formula suggests this is always satisfied so there is no $\alpha$ decay expected for susy nuclei.  This conclusion holds whether one uses the mass formula value or the experimental value for
$m_\alpha$.  The latter value is lower than the former for normal helium and this is expected to be also the case for susy helium since all the nucleons are in their ground state. 

\begin{figure}[htbp]
\begin{center}
\epsfxsize= 3in 
\leavevmode
\epsfbox{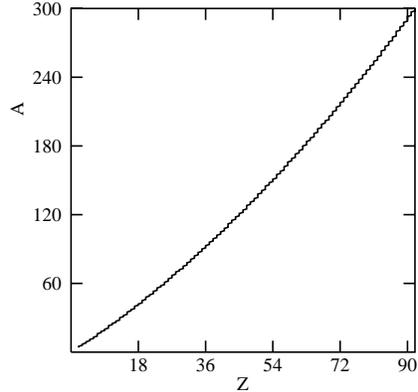}
\end{center}
\caption{Stable susy nuclei in the $Z,A$ plane}
\label{stablenuclei}
\end{figure}

Eq.\,\ref{betaconstraints} implies that the stable nuclei have a unique $Z$ for each $A$
although several isotopes of each element can exist (multiple $A$ for each $Z$).  
\be
     Z = int(\frac{1+A\,\eta}{2})
\ee
where $int(x)$ is the highest integer less than or equal to $x$.  The stable nuclei are shown in fig.\,\ref{stablenuclei}.  Typically, for given $A$, stable nuclei have lower $Z$ than the most stable nuclei in the standard model.  Although $C^{12}$ is stable, the lightest isotope of oxygen has $A=17$ and the lightest isotope of uranium has $A=298$.    

As a corollary of the alpha stability of susy nuclei, the mass formula implies that alpha absorption leads to arbitrary production of high mass nuclei limited only by Coulomb barrier for high $Z$.
An approximately constant energy release of about $80$ MeV occurs with each alpha absorption.
In the susy triple alpha process $3 \alpha \rightarrow C^{12}$, $241$ MeV is released to compare with
$0.47$ MeV in the standard model.  In addition, the intermediate $Be^8$ is stable according to the susy mass formula.  This process would, therefore, occur explosively given an adequate supply of alpha particles such as existed in the early universe.  Since the fine tuning of the triple alpha process is known to be anthropically essential in the standard model, it seems likely that intelligent life would not have arisen if the universe had fallen directly into the susy minimum at the end of the inflationary era.

\section{\bf Conclusion}
\setcounter{equation}{0}

String landscape ideas provide an incentive to investigate the properties of alternative universes that could occur in the future due to vacuum decay or that might exist already in disconnected pieces of the multiverse.  Experimental clues that we might be living in a universe of broken supersymmetry make the investigation of an exactly supersymmetric phase of particular interest.  The theoretical study of this potential universe could also suggest questions about the forces active in our own universe.
In the case at hand, the question arises whether the hard core potential seen in nuclear models and the asymmetry term in the semi-empirical mass formula are solely due to the Pauli Principle.  If so
there would be an extreme preference for sneutron rich nuclei in the susy world \cite{future}.  The alternative discussed here based on the Fermi gas model is that part of the asymmetry term is due to strong-electromagnetic interference or to other effects beside the Pauli Principle.   For simplicity we assume that the
common fermion/sfermion masses in the susy phase are the same as the fermion masses in the broken susy phase.  The effects of alternative assumptions could be investigated.
    
In the standard model, the strong interaction is isospin invariant leading to terms in the nuclear mass formula that depend only on the total number of nucleons $A=N+Z$. The electromagnetic interaction 
leads to a term proportional to the square of the proton number $Z$.  Phenomenologically, an anti-binding asymmetry term 
proportional to $(Z-N)^2/A$ and an alternating $\delta$ term are also needed to fit nuclear masses.
It is tempting to attribute these effective isospin violating effects to the Pauli Principle which
separately affects protons and neutrons.  Although this is a possibility, the
Fermi gas model, which should encode the effects of the Pauli Principle at least for moderate to large nucleon number, only accounts for about one half of the needed asymmetry term.  It is not clear whether such a large departure could be due to a modification of the square well potential.

In this paper, therefore, we have assumed that there is a contribution to the asymmetry term that is not related to the Pauli Principle.  This could come from interference terms involving both strong
and electromagnetic interactions or to new interactions beyond the standard model that distinguish
between up and down quarks.  In this context it is interesting to note the standing puzzle of why
the neutron is heavier than the proton while second and third generation fermions have up-type
partners heavier than down-type.  At present the fermion masses are fixed by a juggling of free parameters. 

If a nucleus is overtaken by a susy bubble, the nucleons and snucleons will become mass degenerate
and the availability of a pair conversion interaction \cite{CP} will result in all particles falling into the ground state with at most two of them remaining fermionic.  This collapse of the Pauli tower will leave only the potential energy terms as given in eq.\,\ref{M3(Z,A)}.  The suggestion, therefore, is that in a susy nucleus the dominant potential binding term proportional to $A$ would be significantly stronger than in our world and the asymmetry term proportional to $(Z-N)^2/A$ would be 
significantly weaker but still non-zero.

The non-zero asymmetry term  
will moderate the sneutron excess found necessary in ref.\,\cite{future} for the stabilization of high $Z$ nuclei.   As a result, if a susy bubble were to engulf an earth-like planet,  common elements up to U(238) would decay down to iridium ($Z=
77$) and below whereas without a Pauli-independent asymmetry term these would decay down to oxygen and below as discussed in ref.\,\cite{future}.  The variational principle calculations of refs.\,\cite{CL} and \cite{CS} imply that atoms and a wide variety of ionic and covalent molecules would then form
suggesting the possibility that advanced life forms could evolve after the susy transition.  

Although no direct experimental confirmation of the existence of a life supporting supersymmetric minimum is possible,  encouraging signs would be the discovery at accelerators of a broken susy with
a singlet extended Higgs structure. Although the exact susy world is not our current universe,  some physics-based thought about its properties should be tolerated as long as string landscape ideas are entertained.

{\bf Acknowledgements}\\
We acknowledge helpful discussions on the issues treated in this paper with G. Goldstein, and S. Liuti.  LC enjoyed the hospitality of Tufts University while this manuscript was being prepared in the summer of 2010.


\begin{thebibliography}{99}
\bibitem{Weinberg} S. Weinberg, {\it Phys. Rev. Le.tt.} {\bf{59}}, 2607 (1987)
\bibitem{Bousso} R. Bousso and J. Polchinski, {\it JHEP} {\bf 0006}: 006 (2000)\\
R. Bousso, hep-th/0610211v3
\bibitem{Susskind} L. Susskind, hep-th /0302219\\
Ben Freivogel, M. Kleban, M.R. Martinez, L. Susskind,
hep-th/0505232
\bibitem{Linde} A. Linde, hep-th/0611043, JCAP 0701:022 (2007)
\bibitem{Giddings} Steven Giddings, hep-th/0303031, {\it Phys. Rev.} D{\bf 68},026006 (2003)
\bibitem{Barr} V. Agrawal, Stephen M. Barr, John F. Donoghue, and D. Seckel,
{\it Phys. Rev.}D{\bf 57}, 5480 (1998)
\bibitem{Donoghue} John F. Donoghue, in {\it Universe or Multiverse}, ed. Bernard Carr, pp231-246,
Cambridge University Press (2007), ArXiv:0710.4080
\bibitem{weakless} R. Harnik, G. Kribs, and G. Perrez, hep-ph/0604027. {\it Phys. Rev.}D{\bf 74}, 035006 (2006)
\bibitem{CW} L. Clavelli and R.E. White III, Problems in a Weakless Universe, hep-ph/0609050
\bibitem{ADS} O. Aharony, S. Gubser, J. Maldacena, H. Ooguri, Y. Oz, {\it Phys. Rep.} 323 (2000)
\bibitem{ISS} K. Intriligator, N. Seiberg, and D. Shih, hep-th/0602239v3, 
JHEP 0604:021 (2006)
\bibitem{future} L. Clavelli, hep-th/0508207, {\it IJMPE} {\bf 15}, 1157 (2006)
\bibitem{Xhiggs} L. Clavelli,Landscape Implications of Extended Higgs Models, ArXiv:0705.1290, Int. Journal of Modern Physics A23, 3509 (2008)\\
Extended Higgs Models and the Transition to Exact Susy,ArXiv:0710.2341,     Proceedings of 15th Annual International Conference on        
Supersymmetry,  Susy07, W. de Boer ed, European Journal of Physics (2008)\\
Metastable Aspects of Singlet Extended Higgs Models, ArXiv:0809:3961
\bibitem{CL} L. Clavelli and T. Lovorn, Ionic Binding in a Susy Background, hep-ph/0611013,  Int. J. of Mod. Phys. A22, 2133 (2007)
\bibitem{CS} L. Clavelli and Sanjoy Sarker, Covalent Molecular Binding in a Susy Background, ArXiv:0811.1022, Int. J. Mod. Phys. A,
\bibitem{CP} L. Clavelli and I. Perevalova,
Electron to Selectron Pair Conversion in a SUSY Bubble,
hep-ph/0409194, Phys. Rev. D71, 055001 (2005)\\
I. Perevalova, University of Alabama PhD dissertation (2010)
\bibitem{BC} P. Biermann and L. Clavelli, to be published
\bibitem{JainSrivastava} S.C. Jain and B.K. Srivastava, {\it Phys. Rev.}{\bf 169}, 833 (1968)\\
S. Aoki, T. Hatsuda, N. Ishii, {\it Prog. Th. Phys.}{\bf 123}, No.1 (2010)

\end{thebibliography}
\end{document}